\documentclass[12pt,letter]{article}

\usepackage{graphicx, epsfig, color}
\usepackage{amsmath,amssymb,hyperref}
\usepackage{subfigure}

\def\to{\rightarrow}

\def\bi{\begin{itemize}}
\def\ei{\end{itemize}}

\def\ta{\tilde a}

\def\sps1ap{SPS1a$^\prime$}
\def\c1p{C1$^\prime$}

\def\tst{\tilde t}

\def\tg{\tilde g}

\def\tw{\widetilde W}
\def\tz{\widetilde Z}
\def\alt{\lesssim}
\def\agt{\gtrsim}
\def\be{\begin{equation}}  
\def\ee{\end{equation}}  
\def\bea{\begin{eqnarray}}  
\def\eea{\end{eqnarray}}  
\def\beas{\begin{eqnarray*}}  
\def\eeas{\end{eqnarray*}}

\newcommand\plb[3]{{\it Phys.\ Lett.\ }{\bf B #1} (#2) #3}

\newcommand{\hepph}[1]{hep-ph/#1}

\begin{document}
\begin{titlepage}
\begin{flushright}
OUHEP-180329\\
UH-511-1291-18
\end{flushright}

\vspace{0.5cm}
\begin{center}
{\Large \bf Is natural higgsino-only dark matter excluded?
}\\ 
\vspace{1.2cm} \renewcommand{\thefootnote}{\fnsymbol{footnote}}
{\large Howard Baer$^1$\footnote[1]{Email: baer@nhn.ou.edu }, 
Vernon Barger$^2$\footnote[2]{Email: barger@pheno.wisc.edu },
Dibyashree Sengupta$^1$\footnote[3]{Email: Dibyashree.Sengupta-1@ou.edu } and
Xerxes Tata$^{3,4}$\footnote[4]{Email: tata@phys.hawaii.edu }
}\\ 
\vspace{1.2cm} \renewcommand{\thefootnote}{\arabic{footnote}} {\it
  $^1$Department of Physics and Astronomy, University of Oklahoma,
  Norman, OK 73019, USA \\ } {\it $^2$Department of Physics,
  University of Wisconsin, Madison, WI 53706, USA \\ } {\it
  $^3$Department of Physics and Astronomy, University of Hawaii,
  Honolulu, HI 96822, USA \\ } {\it $^4$Centre for High Energy Physics,
  Indian Institute of Science, Bangalore 560012, India \\ }
\end{center}

\vspace{0.5cm}
\begin{abstract}
\noindent 
The requirement of electroweak naturalness in supersymmetric (SUSY) models of
particle physics necessitates light higgsinos not too far from the weak scale
characterized by $m_{weak}\sim m_{W,Z,h}\sim 100$ GeV. 
On the other hand, LHC Higgs mass measurements and sparticle mass limits 
point to a SUSY breaking scale in the multi-TeV regime. 
Under such conditions, the lightest SUSY particle is expected to be a 
mainly higgsino-like neutralino with non-negligible gaugino components 
(required by naturalness).
The computed thermal WIMP abundance in natural SUSY models is then found to be
typically a factor 5-20 below its measured value. 
To gain concordance with observations, either an
additional DM particle (the axion is a well-motivated possibility)
must be present or additional non-thermal mechanisms must augment
the neutralino abundance. 
We compare present direct and indirect WIMP detection limits to three
natural SUSY models based on gravity-, anomaly- and mirage-mediation.
We show that the case of natural higgsino-only dark matter
where non-thermal
production mechanisms augment its relic density, is essentially 
excluded by a combination of direct detection constraints
from PandaX-II, LUX and Xenon-1t experiments,
and by bounds from 
Fermi-LAT/MAGIC observations of gamma rays from dwarf spheroidal galaxies.
\vspace*{0.8cm}

\end{abstract}

\end{titlepage}

\section{Introduction}
\label{sec:intro}

Supersymmetric models of particle physics have been under assault from
both collider search experiments and direct and indirect dark matter
detection experiments. From the CERN LHC, the measured value of the
Higgs boson mass $m_h\simeq 125$ GeV\cite{lhc_mhiggs} seems to require
TeV-scale highly mixed top squarks, at least in the framework of the
Minimal Supersymmetric Standard Model, or
MSSM\cite{h125,djouadi,hall,mhiggs}.  Direct searches for superparticles at
LHC have resulted in gluino mass limits $m_{\tg}\agt 2$
TeV\cite{lhc_mgl} and top squark limits $m_{\tst_1}\agt 1$
TeV\cite{lhc_mt1}.  Meanwhile, direct searches for relic WIMP dark
matter by LUX\cite{lux}, PandaX\cite{pandax} and Xe-1-ton\cite{xe1t}
have failed to detect the SUSY WIMP.  Indirect WIMP searches from
Fermi-LAT/MAGIC\cite{fermi}, expecting to detect WIMP-WIMP
annihilation to gamma rays in dwarf spheroidal galaxies, have also
placed strong limits on SUSY WIMPs.  Taken together, direct and
indirect detection limits have eliminated two previously well-regarded
candidates for the nature of SUSY WIMP dark matter.
\begin{enumerate}
\item The well-tempered neutralino (WTN)\cite{wtn}, wherein the bino
  and higgsino
  components were adjusted to comparable values
  so as to obtain the required relic density, predicted
  $\sigma^{SI}(\tz_1 p)\sim 10^{-8}$ pb relatively independently of
  $m_{\tz_1}$. The nucleon-WIMP cross section is roughly independent of the WIMP mass because, for heavier WIMPs, the higgsino component of the WIMP needs to
  be increased to maintain the observed relic density; this increased higgsino
  component then maintains the direct detection cross-section at roughly
  a constant value.
  The light higgsino region is typical of the so-called focus
  point region/hyperbolic branch \cite{fmw} of the mSUGRA/CMSSM model\cite{bbbo}
  and is now solidly excluded\cite{lux,pandax,xe1t,bbs}.

\item The case of wino-like WIMP-only dark matter, which is
  characteristic of anomaly-mediated SUSY breaking models, predicts
  rather large rates for WIMP-WIMP annihilation into $WW$, leading to
  gamma ray production in areas of the universe where increased WIMP
  densities are expected (such as galactic cores and dwarf galaxies).
  Recent limits from Fermi-LAT (at lower $m_{\tz_1}$) and HESS (at
  $m_{\tz_1}\sim$ TeV-scale) have seemingly excluded this possibility
  if one includes Sommerfeld enhancement effects in the annihilation
  cross sections\cite{clps,fr,bbs}. 
\end{enumerate}

Taken all together, the data seem to suggest that weak scale
supersymmetry (WSS)\cite{wss}, if viable, must have
at least strongly coupled superpartners
with soft SUSY breaking parameters $m_{soft}$
in the multi-TeV range rather than at
the weak scale, $m_{weak}\sim m_{W,Z,h}\sim 100$ GeV.
The confrontation of theory with data then seemingly exacerbates what
has become known as the Little Hierarchy problem: why is
$m_{weak}\ll m_{soft}$?  While the introduction of SUSY can solve the
Big Hierarchy problem, avoiding the Higgs mass from blowing up to the
Planck scale while avoiding extreme fine-tuning of parameters, now one
may expect the Higgs boson mass to inflate to the multi-TeV regime
if the heavy superpartners couple directly to the Higgs fields, 
absent
again fine-tuning of SUSY Lagrangian parameters.


The well-known expression for the $Z$-boson mass obtained from the minimization of the (one-loop) scalar potential of the Higgs fields,
\be \frac{m_Z^2}{2} = \frac{m_{H_d}^2 +
\Sigma_d^d -(m_{H_u}^2+\Sigma_u^u)\tan^2\beta}{\tan^2\beta -1} -\mu^2
\simeq  -m_{H_u}^2-\Sigma_u^u(\tst_{1,2})-\mu^2 ,
\label{eq:mzs}
\ee 
serves as a starting point for many discussions of fine-tuning in SUSY
models.  The last (approximate) equality in Eq.~(\ref{eq:mzs}) obtains
for moderate to large values of $\tan\beta$
required by the measured value of $m_h=125$ GeV.
Here, $m_{H_u}^2$ is the weak scale value of the up-Higgs squared soft mass 
and $\mu$ is the Higgs/higgsino mass parameter occuring in the
(SUSY conserving) superpotential. 
The $\Sigma_u^u$ and $\Sigma_d^d$ terms contain an 
assortment of radiative corrections, the largest of which
typically arise from the top squarks:
\be
\Sigma_u^u(\tst_{1,2})=\frac{3}{16\pi^2}F(m_{\tst_{1,2}^2})\left[
f_t^2-g_Z^2\mp\frac{f_t^2A_t^2-8g_Z^2(\frac{1}{4}-\frac{2}{3}x_W)\Delta_t}
{m_{\tst_2}^2-m_{\tst_1}^2}\right],
\label{eq:Sigmat12}
\ee
where $\Delta_t=(m_{\tst_L}^2-m_{\tst_R}^2)/2+m_Z^2\cos 2\beta(\frac{1}{4}-\frac{2}{3}x_W)$ and $x_W\equiv\sin^2\theta_W$. Also, 
$F(m^2)=m^2\left(\log\frac{m^2}{Q^2}-1\right)$ and $g_z^2=(g^2+g^{\prime 2})/8$
and $f_t$ is the top-quark Yukawa coupling.
Expressions for the remaining $\Sigma_u^u$ and $\Sigma_d^d$ are given 
in the Appendix of Ref.~\cite{rns}. 


Requiring no large unexplained cancellations between the various terms
on the right-hand-side of Eq.~(\ref{eq:mzs}) led us to introduce the
{\it electroweak} fine-tuning measure $\Delta_{\rm EW}$\cite{ltr,rns}
defined as the ratio of the magnitude of the maximal contribution on
the right-hand-side (RHS) of Eq.~(\ref{eq:mzs}) to $m_Z^2/2$.  If the
RHS terms in Eq.~(\ref{eq:mzs}) are individually comparable to
$m_Z^2/2$, then no unnatural fine-tuning is required to generate
$m_Z=91.2$ GeV. We advocate the use of $\Delta_{\rm EW}$ for
discussions of fine-tuning because it allows for the possibility that
model parameters traditionally regarded as independent may turn out to
be correlated by the underlying SUSY breaking mechanism, and further,
that the most commonly used fine-tuning measure, $\Delta_{BG}\equiv
max_i|\frac{p_i}{m_Z^2}\frac{\partial m_Z^2}{\partial p_i}|$ where
$p_i$ are fundamental parameters of the theory\cite{bg} reduces to
$\Delta_{\rm EW}$\cite{dew,mt,seige} after appropriate correlations
are incorporated. Ignoring the possibility that model parameters (taken to
be independent) might
turn out to be correlated may lead to prematurely discarding perfectly
viable SUSY models. 
%
 
The most important implications of low electroweak fine-tuning (which we take
to be $\Delta_{\rm EW}\alt 30$) \footnote{ The onset of fine-tuning
  for $\Delta_{\rm EW}\agt 30$ is visually displayed in
  Ref.~\cite{upper}.} are the following.
\begin{enumerate}
\item $|\mu |\sim 100-300$ GeV~\cite{Chan:1997bi,bbh} (the lighter the
  better) where the higgsino mass $\sim \mu \agt 100$ GeV 
  to accommodate LEP2 limits from chargino pair production
  searches.\footnote{Here, we have implicitly assumed that $\mu$ is
    independent of the soft-SUSY-breaking parameters (for a very
compelling model, see {\it e.g.} Ref. \cite{radpq}), and further that
    it makes the dominant contribution to the higgsino mass.}
\item $m_{H_u}^2$ is driven radiatively from its high scale value to small
negative values, comparable to $-m_Z^2$, at the weak scale~\cite{ltr,rns}.
\item The top squark contributions to the radiative corrections
$\Sigma_u^u(\tst_{1,2})$ are minimized for TeV-scale highly mixed top
squarks~\cite{ltr}. This latter condition also lifts the Higgs mass to
$m_h\sim 125$ GeV. For $\Delta_{\rm EW}\alt 30$, the lighter top
squarks are bounded by $m_{\tst_1}\alt 3$ TeV~\cite{rns,upper}.
\item The gluino mass, which feeds into the stop masses at one-loop 
and hence into the scalar potential at two-loop order,
is bounded by $m_{\tg}\alt 6$ TeV~\cite{rns,upper}.
\end{enumerate}

We will collectively call SUSY models for which $\Delta_{EW}<30$ {\it
  natural} SUSY models.  In the present paper, we examine expectations
for SUSY WIMP dark matter from three different, well-motivated classes
of natural SUSY models that lead to qualitatively different patterns
of gaugino and higgsino masses which in turn determines the nature of the SUSY
WIMP.
\begin{enumerate}
\item Gravity-mediated SUSY breaking models, as exemplified by the
  two-extra-parameter {\it natural} non-universal Higgs model\cite{nuhm2} 
(nNUHM2)
  with parameter space given by\footnote{Models such as mSUGRA/CMSSM
    where all soft scalar masses are set to $m_0$ are no longer
    natural for $m_h=125$ GeV while respecting LHC sparticle mass
    limits\cite{firstsugra,dew,seige}. 
    Historically, the common mass $m_0$ originated
    in the assumption of a flat K\"ahler potential in minimal
    supergravity models.  The CMSSM requirement that
    $m_{H_u}=m_{H_d}=m_0$ appears ad-hoc and artificial given that the
    Higgs multiplets belong to different GUT representations from the
    matter scalars. The NUHM2 model rectifies the artificial
    degeneracy requirement and then allows for naturalness whilst
    respecting LHC Higgs mass and sparticle search constraints. Indeed
    the NUHM3 generalization where third generation scalars are
    treated differently from scalars of the first two generations has
    also been examined in the literature.}
  \be
  m_0,\ m_{1/2},\ A_0,\ \tan\beta, \mu,\ m_A\ \ \ (nNUHM2).
  \ee
  For NUHM2, because of the gaugino mass unification assumption,
  one expects weak scale gaugino masses in the ratio
  $M_1:M_2:M_3\sim 1:2:7$.
  
\item A phenomenological generalization of the well-studied
  anomaly-mediated SUSY breaking model, the natural (generalized)
  anomaly-mediated SUSY breaking model\cite{namsb} (nAMSB) with
  parameter space given by,
  \be
m_{3/2},\ m_0(1,2)(bulk),\ m_0(3)(bulk),\ A_0(bulk),\ \tan\beta,\ \mu ,\ m_A
  \ \ \ (nAMSB),
  \ee
  where in addition to AMSB contributions to soft terms\cite{amsb}, we introduce
  several {\it bulk} induced soft terms to render sleptons
  non-tachyonic ($m_0(bulk)$) and to render the model natural
  ($m_{H_u}(bulk)$, $m_{H_d}(bulk)$ and $A_0(bulk)$) whilst respecting
  LHC data. As in NUHM2, we trade the high scale parameter freedom of
  $m_{H_u}(bulk)$ and $m_{H_d}(bulk)$ for the more convenient weak
  scale parameters $\mu$ and $m_A$. For nAMSB, one expects weak scale
  gaugino masses in the ratio $M_1:M_2:M_3\sim 3:1:8$ but now with
  $\mu< M(gauginos)$ so that a higgsino-like neutralino (mixed with some
  wino component) is the lightest SUSy particle (LSP) instead of the neutral wino.

\item Natural generalized mirage mediation model (nGMM) where 
  gravity- and anomaly-mediated contributions to soft SUSY breaking terms are
  comparable. The nGMM mass pattern is expected to emerge from several
  well-motivated superstring models\cite{mirage}. The parameter space
  is given by\cite{gmm,mini},
  \be
  \alpha,\ m_{3/2},\ c_{m},\ c_{m3},\ a_3,\ \tan\beta,\ \mu,\ m_A\ \ \ (nGMM),
  \ee
  where $\alpha$ parametrizes the relative gravity- to anomaly-
  mediation and $c_m$ and $c_{m3}$ co-efficients are a continuous
  generalization of formerly discrete parameters involving modular
  weights of the relevant fields and $a_3$ is a continuous
  generalization of the formerly discrete trilinear gravity-mediated
  $A$ term. For nGMM models, one expects the weak scale gaugino masses
  with $M_1<M_2<M_3$ but with compressed spectra (depending on the
  value of $\alpha$ since the scale of mirage unification
  $\mu_{mir}=m_{GUT}e^{-8\pi^2/\alpha}$) since they appear to unify at some
  intermediate scale rather than $m_{GUT}\simeq 2\times 10^{16}$ GeV.
\end{enumerate}

\section{Dark matter relic density in natural SUSY}

For natural SUSY models, we see from Eq.~(\ref{eq:mzs}) that
naturalness requires $|\mu|$ and $-m_{H_u}^2$ to be not too far above
$m_Z^2$, but imposes only  loop-suppressed restrictions on other soft
SUSY breaking parameters. Hence, one expects the LSP to be {\it dominantly} higgsino-like, but with a
non-negligible gaugino component (lest $\Sigma_u^u(m_{\tw_2})$ becomes
large for too large wino masses). The first question then is: do the
natural SUSY models produce the measured relic abundance of dark
matter in the universe given by $\Omega_{DM}h^2\equiv
\frac{\rho_{DM}}{\rho_c} h^2$ where $\rho_c$ is the critical closure
density of dark matter and $h$ is the scaled Hubble parameter. Of
course, since higgsinos annihilate with full gauge strength in the
early Universe, we do not expect that the relic density of {\em
  thermally produced}, light higgsinos to saturate the observed relic
density, but it is nonetheless instructive to examine the expectations
for the thermal relic
density in well-motivated natural SUSY models.

To answer this question, we next compute the thermally-produced relic
density for the various SUSY models introduced in
Sec.~\ref{sec:intro}.  We use the computer code Isajet 7.88 to compute
sparticle mass spectra for the nNUHM2, nAMSB and nGMM
models\cite{isajet}.  For nNUHM2 and nAMSB models,
we have performed a broad random scan  as well as an additional
focused scan (over the parameter ranges shown in parenthesis below)
in an attempt to further
zero in on the natural SUSY region of the parameter space, while for the
nGMM model, our scan is already quite focussed.
For the NUHM2 model we scan over the parameter range :
\bea
m_0&:&\ 0-10\ {\rm TeV},\nonumber\\
m_{1/2}&:&\ 0.5-3\ {\rm TeV}, (0.7-2\ {\rm TeV}),\nonumber\\
A_0&:&\ -20\ \to\ +20\ {\rm TeV},((-1\ \to\ -3)m_0),\\
\tan\beta &:&\ 4-58,\nonumber\\
\mu &:&\ 100-500\ {\rm GeV}, (100-360\ {\rm GeV}),\nonumber\\
m_A &:&\ 0.25-10\ {\rm TeV}.\nonumber
\eea
For nAMSB model, we scan over
\bea
m_{3/2}&:&\ 80-1000\ {\rm TeV}, (80-300\ {\rm TeV}),\nonumber\\
m_0(3)&:&\ 1-10\ {\rm TeV},\nonumber\\
m_0(1,2)&:& m_0(3) - 20\ {\rm TeV},\nonumber\\
A_0&:&\ -20\ \to\ +20\ {\rm TeV},((+0.5\ \to\ +2)m_0(3)),\\
\tan\beta &:&\ 4-58, \nonumber\\
\mu &:&\ 100-500\ {\rm GeV},(100-350\ {\rm GeV}),\nonumber\\
m_A &:&\ 0.25-10\ {\rm TeV}.\nonumber
\eea
For the nGMM model, we scan over
\bea
\alpha &:&\ 2-40\ \nonumber\\
m_{3/2}&:&\ 3-65\ {\rm TeV}\nonumber\\
c_m&=&\ (16\pi^2/\alpha )^2\ \nonumber\\
c_{m3}&:&\ 1-{\rm min}[40,(c_m/4)], \\
a_3&:&\ 1-12, \nonumber\\
\tan\beta &:&\ 4-58, \nonumber\\
\mu &:&\ 100-360\ {\rm GeV},\nonumber\\
m_A &:&\ 0.3-10\ {\rm TeV}.\nonumber
\eea
For each solution, we require the light
Higgs boson $m_h:\ 122-128$ GeV (allowing for $\pm 3$ GeV error in the
Isajet $m_h$ calculation).\footnote{In our previous studies of naturalness
  we had used a $\pm 2$~GeV window on $m_h$. We have checked that our
  conclusions are quite insensitive to this wider and more conservative
  window.} To enforce naturalness, we require of each
solution $\Delta_{EW}<30$.  We also require $m_{\tg}>2$ TeV and
$m_{\tst_1}>1$ TeV in accord with LHC sparticle search limits.

\begin{figure}[tbp]
\begin{center}
\includegraphics[height=0.4\textheight]{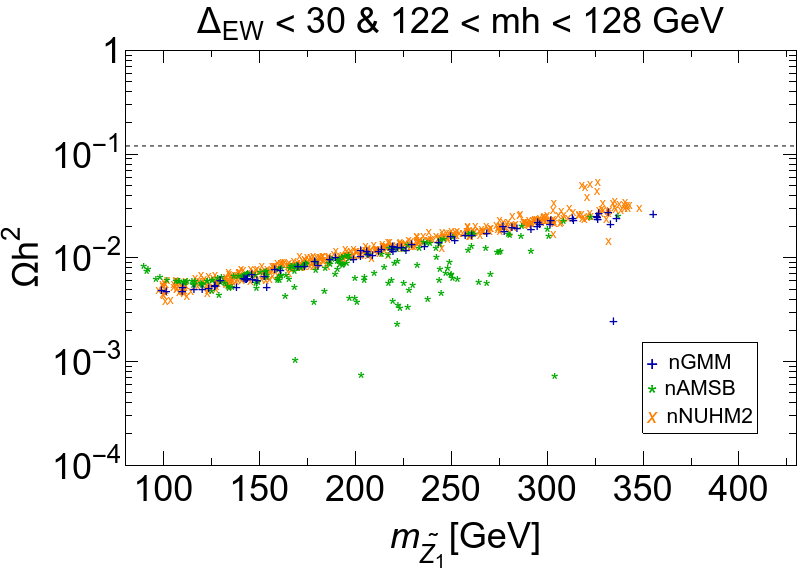}
\caption{Plot of points in the $\Omega_{\tz_1}h^2$ vs. $m_{\tz_1}$ plane 
from a scan over the natural NUHM2, nGMM and nAMSB model parameter space.
The dashed line shows the measured value.
\label{fig:Oh2}}
\end{center}
\end{figure}

The results of our calculations of the thermal LSP relic density $\Omega_{\tz_1}^{TP}h^2$ (using the Isajet subcode IsaReD\cite{isared})
are shown versus $m_{\tz_1}$ in
Fig. \ref{fig:Oh2} for the three natural SUSY models. We plot points
from our scan that yield $\Delta_{\rm EW} \le 30$ and also satisfy the
Higgs boson mass and LHC sparticle mass constraints as blue pluses
(nGMM model), green stars (nAMSB model) and yellow crosses (nNUHM2
model).  We see first that $m_{\tz_1}$ is bounded from below by
$m_{\tz_1}\agt 100$~GeV due to LEP2 limits on $m_{\tw_1}\agt 100$ GeV
(which we set as the lower limit on the $\mu$ parameter scan). Also,
$m_{\tz_1}$ is bounded from above by $m_{\tz_1}\alt 350$ GeV from the
naturalness constraint, $\Delta_{EW}<30$. For the lower range of
$m_{\tz_1}$ values, then $\Omega_{\tz_1}^{TP}h^2$ is typically a
factor $\sim 20$ below the measured value $\Omega_{\rm CDM} = 0.1199 \pm 0.0022$
\cite{planck}
while for the high range of
$m_{\tz_1}$ then the calculated relic abundance is about a factor
$\sim 4$ below the measured result. The range of under-abundance just mentioned
applies to all three models with the possible exception of nAMSB where
some of the green stars lie at even lower $\Omega_{\tz_1}^{TP}h^2$
values. The reason for this is that in nAMSB models, for a lower range
of $m_{3/2}$ values then the wino can range down to $M_2:200-300$ GeV
so that for this model the $\tz_1$ can be mixed higgsino-wino variety:
then the neutralino annihilation rate in the universe is enhanced even
beyond the higgsino-like case leading to even lower relic
density. Thus, natural SUSY models typically predict an
under-abundance of {\em thermally produced neutralinos} in standard
Big Bang cosmology by a factor $\sim 5-25$. Other mechanisms are
required to bring the expected DM abundance into accord
with data.

Two well-motivated classes of mechanisms have been proposed to bring
thermally-produced under-abundance of neutralinos into accord with the
measured dark matter abundance. In the first class, the dark matter is
{\it multi-component} with thermal higgsinos comprising only a fraction of
the observed dark matter, with the remainder consisting of other
particle(s).  The axion 
is perhaps the best-motivated candidates for the remainder of the dark
matter (for a review, see {\it e.g.} Ref. \cite{kimrev}). 
In the second class of models, the dark matter is {\em all
  neutralinos}, with a non-thermal component from late decays (to
neutralinos) of heavy particles making up the balance of the observed
relic density. We will see below that if the neutralino is dominantly
the higgsino of natural SUSY, the second class of models is 
essentially ruled out by the data.

\subsection{Mixed axion/WIMP dark matter}

As mentioned, one possibility is that the total WIMP abundance does
not saturate the measured relic density but that, like visible matter,
the dark matter is comprised of several particles.  A very natural choice
for a second dark matter particle is the QCD axion which also
seems to be required to solve the strong CP
problem in QCD.  In a supersymmetric context, then the axion should
occur as but one element of an axion superfield which would also
necessarily contain a spin-0 $R$-parity even saxion field $s$ and a
spin-${1\over 2}$ $R$-parity-odd axino field $\ta$.  Both saxion and
axino are expected to gain masses of order the gravitino mass
$m_{3/2}$ in supergravity models\cite{lukas}. 
In SUSY axion models, the axions can
be produced non-thermally via 1. vacuum misalignment, 2. thermally, and also
3. non-thermally via (late time) saxion decay $s\to aa$. The latter two
may lead to relativistic axions whose population is limited by strict
bounds on the effective number of relativistic degrees of freedom
$N_{\rm eff}=3.15\pm 0.23$ derived from fits to CMB and other
cosmological data \cite{planck}.  Axinos can be thermally produced in
the early universe and then augment the WIMP abundance via decays
after thermal WIMP freeze-out. Saxions can be produced both thermally
and non-thermally and then decay to SM particles (resulting in {\it
  entropy dilution} of all relics from their value at the time of
decay), SUSY particles (which augment the WIMP abundance) or to axions
as mentioned above. WIMPs can be produced thermally or non-thermally
via axino, saxion or gravitino decay. The resultant mixed axion-WIMP
abundance has been evaluated by solving eight-coupled Boltzmann
equations\cite{boltz}. For low values of the axion decay constant, $f_a\alt
10^{11}$ GeV, the WIMP abundance is its thermal value since axinos and
saxions tend to decay before WIMP freeze-out. If $f_a\agt 10^{11}$
GeV, then post-freeze out saxion and axino decays may augment the WIMP
abundance.  The exact rates also depend on the underlying SUSY axion
model assumed (KSVZ or DFSZ), as well as on other parameters such as
$m_{\ta}$, $m_s$, $\theta_s$, $m_{3/2}$ and the SUSY particle mass
spectrum (which influences the saxion and axino decay branching
fractions)\cite{boltz}.

The upshot is that the expected rates for direct and indirect WIMP
detection now depend on the fractional WIMP abundance denoted by
$\xi=\Omega_{\tz_1}h^2/0.12 < 1$ since now there are fewer target
WIMPs compared to the WIMP-only hypothesis for dark matter.  For
spin-independent (SI), spin-dependent (SD) detection rates, and also
the neutrino detection rate at IceCube, the target event rates must be
scaled by a factor $\xi$\footnote{Assuming the WIMP density in the sun
  is in equilibrium, the WIMP annihilation rate used to determine the
  (bound on the) spin-dependent cross section at IceCube is fixed by
  the WIMP {\em capture rate} which scales linearly as $\xi$, and has no
  further dependence on the WIMP annihilation cross section.} while
for indirect WIMP detection (IDD) via WIMP-WIMP annihilation into
gamma-rays or particle-antiparticle pairs, the event rates must be
scaled by a factor $\xi^2$. For mixed axion/WIMP dark matter, we will
assume $\xi=\Omega_{\tz_1}^{TP}h^2/0.12$ which is usually the lower
bound on $\xi$. For special cases at high $f_a$, bosonic collective
motion (BCM) produces a large saxion abundance in the early universe.
If parameters are adjusted properly (the $saa$ coupling is tiny or
zero to avoid relativistic axion production and $m_s<2m_{\tz_1}$ so
$s$ decays only to SM particles) then it is possible to have large
entropy dilution of all relics\cite{largefa} and even lower $\xi$
values; this seems rather contrived, and we will ignore this
possibility in this paper.

\subsection{Non-thermally produced WIMP-only dark matter}

Another option is to assume WIMP-only dark matter where the
additional WIMP abundance is assumed to arise from non-thermal
processes. The prototypical non-thermal WIMP production process occurs
from light modulus field $\phi$ production in the early universe via
the BCM (which also occurs for saxion and cold axion production).  If
the modulus field (of mass $m_\phi$) then decays after WIMP freeze-out
but before the onset of BBN, then it may augment the
thermally-produced abundance to gain accord with the measured density
of dark matter.  This mechanism was originally suggested by Moroi and
Randall \cite{mr} to account for how wino-like LSPs from AMSB models
could account for the observed dark matter. It was later emphasized by
Gondolo and Gelmini\cite{gg} that the measured relic density could be
achieved for any value of
$\Omega_{\tz_1}^{TP}h^2>10^{-5}(100\ GeV/m_{\tz_1})$ by adjusting just
two parameters: $b/m_{\phi}$ and $T_{R2}$ where $b$ is the number of
neutralinos produced per $\phi$ decay and $T_{R2}$ is the (second)
reheat temperature arising from $\phi$ decay.  This reheating temperature is
related to the $\phi$ field energy density as $T_{R2}\sim
\rho_\phi^{-1/4}$. Non-thermal WIMP production has also been recently
invoked to reconcile an underproduced WIMP relic density with measured
value in string-motivated models with a wino-like
LSP\cite{kane,kuver,bhaskar}. For the case of natural WIMP-only dark
matter, we will assume the thermal and non-thermal relic density
contributions sum to the measured dark matter density so that $\xi =1$
for this case.

\section{Bounds on natural SUSY WIMPs from direct and indirect WIMP searches}
\label{sec:DD_IDD}

\subsection{Direct WIMP detection bounds}

\begin{figure}[tbp]
\begin{center}
\includegraphics[height=0.45\textheight]{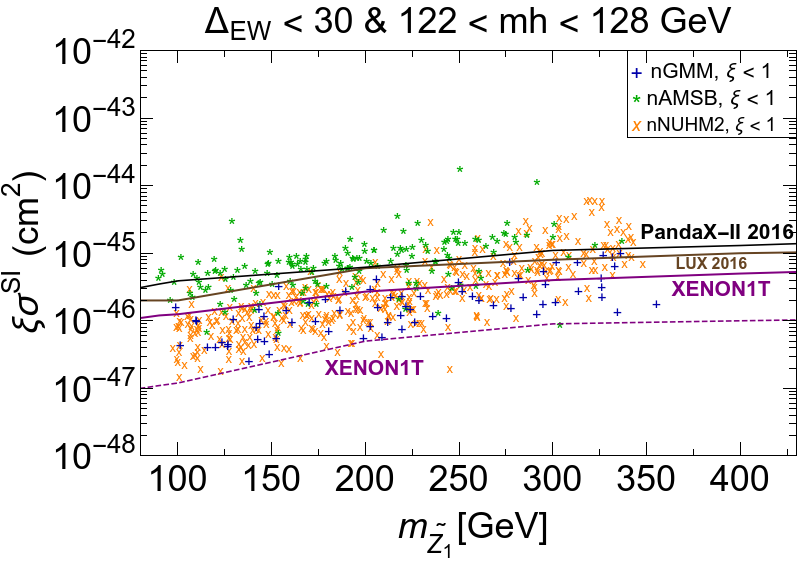}
\includegraphics[height=0.45\textheight]{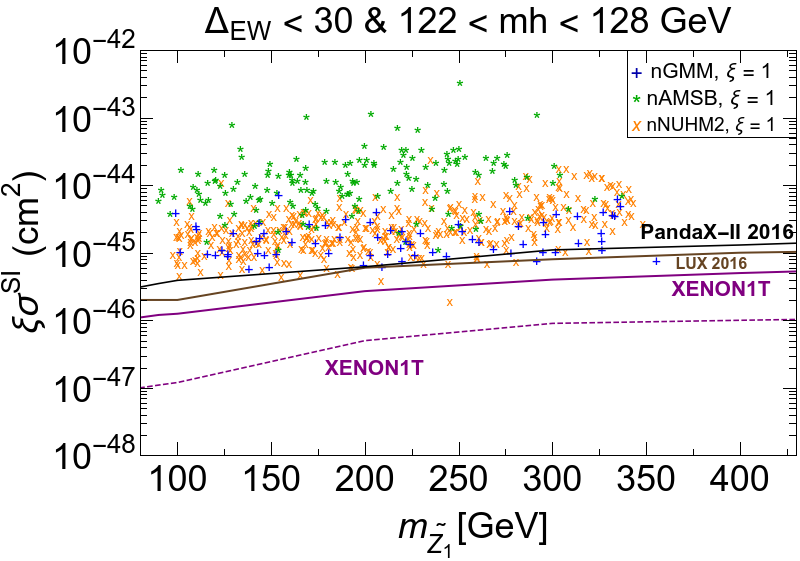}
\caption{Plot of points in the $\sigma^{SI}(\tz_1 p)$ vs. $m_{\tz_1}$ plane 
from a scan over the natural NUHM2, nGMM and nAMSB model parameter space
for {\it a}) $\xi < 1$, assuming the neutralino relic density is
given by its thermal value, and {\it b}) $\xi =1$. 
\label{fig:si}}
\end{center}
\end{figure}

In Fig. \ref{fig:si}, we show the value of $\xi\sigma^{SI}(\tz_1 p)$
vs. $m_{\tz_1}$ for {\it a}) the case with $\xi
=\Omega_{\tz_1}^{TP}h^2/0.12 < 1$ (corresponding to mixed axion/WIMP
DM with no non-thermal WIMP production or dilution) while in frame
{\it b}) we show the case with natural WIMP-only DM and $\xi =1$. 
We use the Isajet subcode IsaReS\cite{isares} for our direct and indirect 
relic scattering calculations.
In both frames, we also plot the current SI DD bounds from LUX, PandaX
and Xe-1ton (solid curves), along with a future projected bound from
Xe-1ton (dashed).  From frame {\it a}), we see that present bounds
already exclude many natural SUSY model points even with $\xi < 1$, if
we assume that the neutralino relic density is given by its thermal
value. Especially, a large fraction of nAMSB model points are
excluded.  This is because in nAMSB the winos can be relatively light
compared to $m_{\tg}$ and the $h\tz_1\tz_1$ coupling occurs as a product of
gaugino times higgsino components (see Eq.~(8.117) of
Ref. \cite{wss}). The enhanced $\tz_1 p$ scattering rate for nAMSB
more than compensates for the somewhat diminished relic abundance. For
the nNUHM2 and nGMM models, the major portion of model points survive
the current SI DD bounds. But future ton-scale noble liquid search
experiments will cover the remainder of parameter space, assuming that
the neutralino relic density is not diluted from its thermal value by
entropy injection in the early Universe. 

In frame {\it b}), for WIMP-only DM with $\xi =1$, then we see that
current bounds exclude almost every point of all three models. A
single point from the scan with $m_{\tz_1}\sim 250$ GeV has survived.
The surviving point lies within the future reach of ton-scale noble
liquid detectors. Thus, it appears from this plot alone that natural
WIMP-only DM appears to be essentially  excluded (but for one nNUHM2 point
which, we have checked, has gaugino masses close to their naturalness
upper limit, and hence a reduced gaugino content and correspondingly reduced
neutralino coupling to $h$).

\begin{figure}[tbp]
\begin{center}
\includegraphics[height=0.45\textheight]{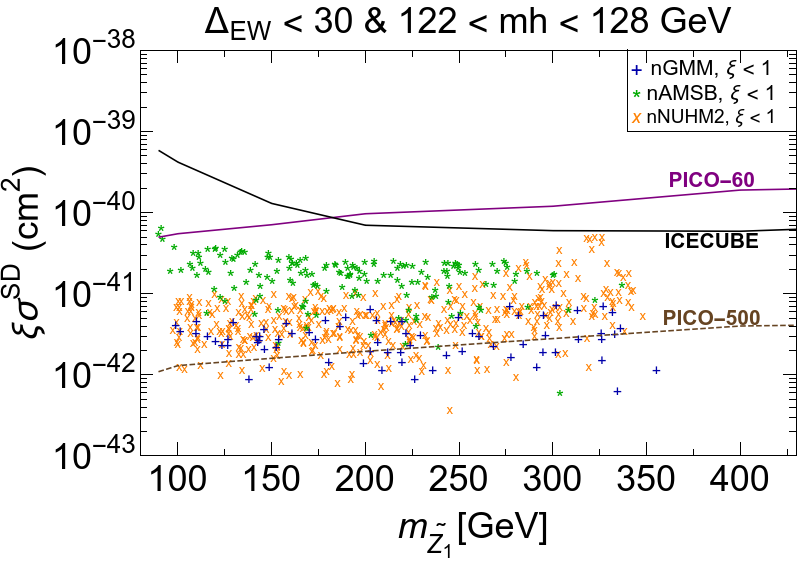}
\includegraphics[height=0.45\textheight]{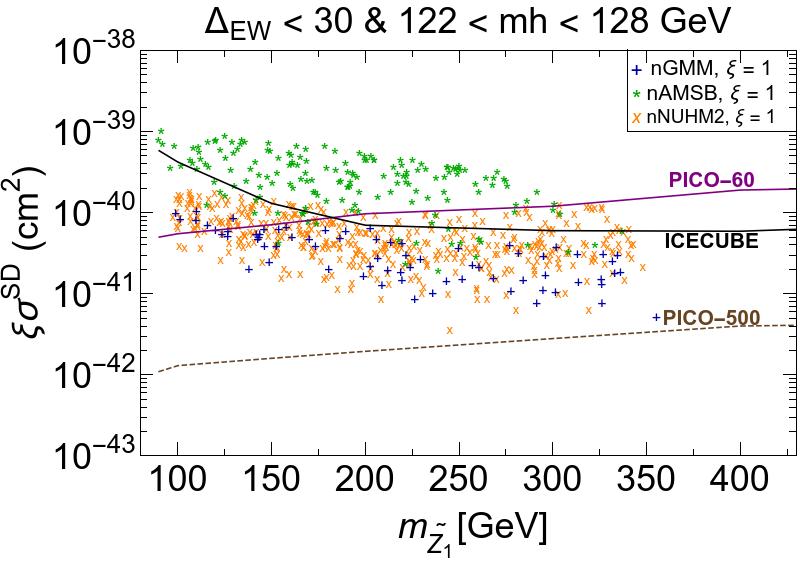}
\caption{Plot of points in the $\sigma^{SD}(\tz_1 p)$ vs. $m_{\tz_1}$ plane 
  from  scans over the parameter space of the
 the natural NUHM2, nGMM and nAMSB models
for {\it a}) $\xi < 1$, assuming the neutralino relic density is
given by its thermal value, and {\it b}) $\xi =1$.
\label{fig:sd}}
\end{center}
\end{figure}

In Fig. \ref{fig:sd}, we show $\xi\sigma^{SD}(\tz_1 p)$
vs. $m_{\tz_1}$. Again, in frame {\it a})
we take $\xi=\Omega_{\tz_1}^{TP}h^2/0.12 < 1$
while in {\it b}) we show the
natural WIMP-only case with $\xi =1$. We also show the current SD
limits from the PICO-60 experiment\cite{pico60} and from
IceCube\cite{icecube} (the latter assuming dominant WIMP annihilation
within the solar core into $WW$ final states). From frame {\it a}), we
see that, save for a few points around $m_{\tz_1}\simeq 100$ GeV, all
points avoid the present SD DD bounds.
We also see that the bulk of natural SUSY points will be probed by PICO-500 \cite{pico500} (subject to the caveats mentioned above)
although some points might still elude SD detection.

In frame {\it b}), we show the $\xi =1$ case for natural WIMP-only DM.
In this case, we see that a combination of PICO-60 and IceCube have
already ruled out a significant fraction of natural SUSY model
points. The projected reach of PICO-500 should probe the remaining
possibilities.

\subsection{Indirect WIMP detection bounds}

\begin{figure}[tbp]
\begin{center}
\includegraphics[height=0.4\textheight]{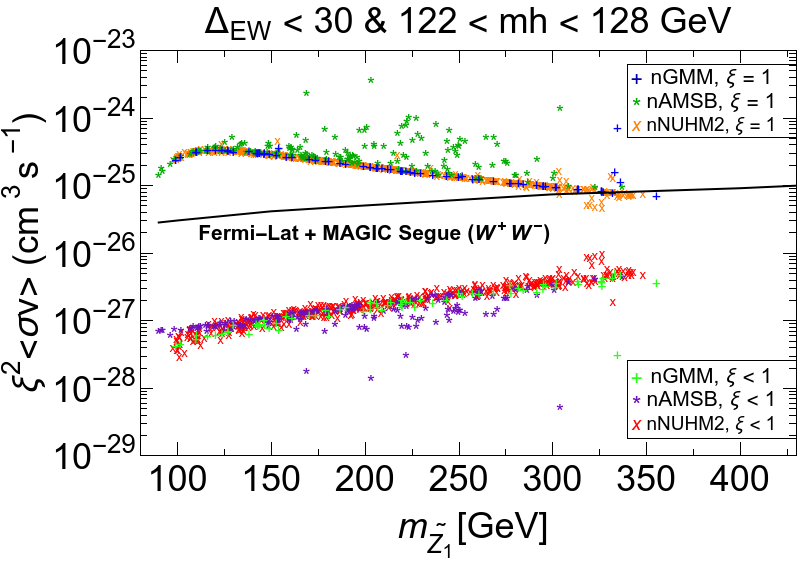}
\caption{The scaled values $\xi^2\langle\sigma v\rangle$
  from  scans over the parameter space of the
 the natural NUHM2, nGMM and nAMSB models
for  $\xi < 1$, assuming the neutralino relic density is
given by its thermal value (lower set), and  $\xi =1$ (upper set).
  \label{fig:sigv}}
\end{center}
\end{figure}

In Fig. \ref{fig:sigv}, we show the quantity $\xi^2\langle\sigma
v\rangle$, the thermally averaged WIMP-WIMP annihilation cross section
times velocity, evaluated as $v\to 0$, scaled by the square of the
depleted relic abundance, vs. $m_{\tz_1}$. In this figure, the mixed
axion/WIMP dark matter points with $\xi\ll 1$ (lower set of points),
again assuming the thermal neutralino relic density is close to its
real value, are neatly separated from the $\xi =1$ points for
WIMP-only dark matter (upper set of points). We also show the present
bounds from the combined Fermi-LAT and MAGIC collaborations derived from
observations of gamma rays from dwarf spheroidal galaxies. Corresponding
limits from HESS
are relevant only for higher, unnatural values of $m_{\tz_1}$, and not shown
in the figure. We see that {\it all} of the mixed axion/WIMP dark matter
points fall well below the experimental bounds.  However, we also see
that all the natural WIMP-only points with $\xi =1$ points are
excluded by present bounds save for a few points with $m_{\tz_1} >300$
GeV.  These $m_{\tz_1}>300$ GeV points are excluded by
the SI DD bounds from Fig. \ref{fig:si}. Likewise, the lone nNUHM2
point with $m_{\tz_1}\sim 250$ GeV is excluded by the IDD bounds with
$\xi =1$.

\section{Concluding Remarks}
\label{sec:conclude}

In this paper we have examined the direct- and indirect- WIMP
detection rates\footnote{For related recent work on AMS-02 bounds using
$\bar{p}$ rates on non-natural SUSY models, 
see Ref's \cite{Giesen:2015ufa} and \cite{Cuoco:2017iax}. 
For recent work on direct, indirect and collider
constraints on thermal-only SUSY WIMPs, 
see {\it e.g.} \cite{Bramante:2015una}.
For general constraints on higgsino dark matter, see
Ref. \cite{Aparicio:2015sda}.} 
for three different natural SUSY models with very
different gaugino spectra: nNUHM2, nAMSB and nGMM. The three models
all have higgsino-like LSPs but qualitatively different and
non-negligible gaugino components.  They have suppressed values of
{\em thermally produced} neutralino relic abundances -- lower than the
measured abundance of CDM by factors ranging from 5-25.  For the three
models, we have examined their WIMP SI- and SD- direct detection rates
and also their indirect detection rates for two different
possibilities: 1. mixed axion-WIMP dark matter where only a fraction
$\xi$, determined by the thermal neutralino relic abundance, is
assumed to be due to WIMPs, while the remainder is axions, and 2. the case
of WIMP-only dark matter where the thermal relic abundance is
supplemented by non-thermal production from processes like modulus
field decay in the early universe. In this second case, then we take the
fractional WIMP abundance $\xi =1$.

From our scans of the parameter space of natural SUSY models, we find
that models where the WIMP relic density (taken to be its thermal
value) forms just $\sim$5-20\% of the measured CDM density comfortably
survive constraints from LHC as well as those from direct and indirect
searches. Direct searches at ton-sized detectors (Xenon-nT or LZ) will probe
the entire natural SUSY parameter space, {\em assuming that the relic
  abundance is given by its thermal expectation.} In this case, future
experiments such as PICO-500 -- designed to measure the spin-dependent
neutralino-nucleon scattering -- will also probe a large part (but not
all) of the parameter space. Otherwise, future colliders such as an
electron-positron collider with $\sqrt{s} \ge 500-600$~GeV \cite{ee}, or a
high energy $pp$ collider operating at $\sqrt{s} \sim
27-33$~TeV\cite{jamie} will be necessary for a definitive probe of the
natural SUSY scenario with multi-component dark matter.

The situation for natural SUSY models where the neutral higgsino-like
WIMP saturates the observed relic density is qualitatively
different. These scenarios are essentially excluded both by bounds
from direct detection experiments as well as by {\em independent}
bounds from Fermi-Lat + Magic observations of high energy gamma rays
from dwarf galaxies.  More correctly, while a few points from our
scans survive the indirect searches, these are excluded by direct
detection, and vice-versa. Such models would also be decisively probed
by spin-dependent direct-detection at PICO-500.
%
%

Thus, the answer to the question posed in the title is: yes, it
appears the case of natural higgsino-like-WIMP-only dark matter is indeed
excluded. Unnatural higgsino-like WIMP dark matter can still survive
as detailed in Ref. \cite{rosz,kowalska} although these models would
have a difficult time explaining why it is that the weak scale is a
mere 100 GeV instead of lying in the multi-TeV range. 
Another possibility is to have models with non-universal gaugino masses
where $M_3>2$ TeV to satisfy LHC gluino mass bounds but where 
$M_1\sim 50-150$ GeV with $|M_1 |< |\mu |$. This case, explored with
running non-universal gaugino masses in Ref.~\cite{maren} 
and in the pMSSM context in the first of Ref.~\cite{caron}, has a mainly 
bino-like LSP while still satisfying naturalness bounds. 
It is unclear as to the origin of the rather large mass gap between
bino and gluino.

As a whole, our results seem to bolster the case for a 
second dark matter particle such as the axion. 
While the remainder of the dark matter could be in the
hidden sector, the axion is a very well motivated candidate 
which may well constitute the bulk of dark matter in
our Universe. Prospects for the complementary axion searches in SUSY
axion models have been examined in Ref. \cite{bbs_ax}.

\section*{Acknowledgments}

This work was supported in part by the US Department of Energy, Office
of High Energy Physics. XT thanks the Centre for High Energy Physics,
Indian Institute of Science for their hospitality during the course of
this work and also the Infosys Foundation for financial support that
made his visit to Bangalore possible.

%

%
\end{document}